\documentclass{amsart}

\usepackage{algorithm}
\usepackage{algorithmic}
\floatname{algorithm}{Procedure}
\usepackage{hyperref}
\hypersetup{unicode}
\usepackage{amsmath}

\numberwithin{algorithm}{section}
\usepackage{graphicx}
\newcommand{\QS}[1]{\lq{#1}\rq}
\newcommand{\Q}[1]{{\QS{#1}}}
\newcommand{\B}[1]{({#1})}
\newcommand{\abs}[1]{\lvert#1\rvert}
\newcommand{\nest}[1]{\left({#1}\right)}
\newcommand{\f}[2]{#1\nest{{#2}}}

\newcommand{\Empty}[0]{\oslash}
\newcommand{\Rset}[0]{\mathbf R}

\newcommand{\dom}[1]{\f{\mathrm{dom}}{#1}}

\newcommand{\gw}[1]{\|{#1}\|}

\newcommand{\suchthat}[0]{\lvert}

\begin{document}
\title{Solution Bounds for a Hypothetical Polynomial Time Aproximation
Algorithm for the TSP}
\author{A. G. Yaneff}
\begin{abstract}
Bounds for the optimal tour length for a hypothetical TSP algorithm are
derived. 
\end{abstract}
\maketitle
The present is an exemplification of part of \cite{CORR209027},
and contains some points made
in \cite{yaneff1} in reference to the TSP.

Consider the hypothetical (and restricted) problem:
\begin{quote}
Given\footnote{The pair $\nest{E:V\times V,
\nest{d:V\times V\rightarrow \Rset\suchthat \dom d = E}}$ is used
instead of the standard graph notation $G=(V_E,E)$.
Connectivity can be represented by a value 
$\beta < \infty$, outside solution value bounds, chosen in
complement to the optimisation objective.} $\nest{E,d}$,
construct such $S\subseteq E$,
such that (i)\footnote{$\gw X = \sum_{e:X}\f d e$} $1\leq \frac{\gw S}{\gw {S_0}}\leq \xi$;
(ii) $S$ is of structure as that of $S_0$; and
(iii) $\abs S = \abs {S_0}$, where $\gw{S_0}$ is an
extremal\footnote{There are exactly two extremal values
- a minimum $\gw{S_0}$ and a maximum $\gw{S_1}$} value.
\end{quote}
The ratio $1\leq \frac{\gw S}{\gw {S_0}}\leq \xi$ is examined.
For simplicity\footnote{Within the context of these notes, the structure
of $S$ need not be considered.} (ii) and (iii) are presumed to be
satisfied by the $PTA$, which constructs an approximate 
solution for the TSP. In a disciplined-like way this is
specified as:
\[
\lbrace S = \Empty \rbrace\,\,
\mathbf{PTA}\,\,
 \lbrace S = S_{\nest {\abs{S_0}}} \land 
 S \subseteq E \land \abs S = \abs{S_0}\land 1
 \leq\frac{\gw {S_{\nest {\abs{S_0}}}}}{\gw{S_0}}\leq \xi\rbrace
\]
\par
$\xi$ as bound
on the performance\footnote{Bracketed subscripts used on solution values
 to avoid confusion with extremums.}
ratio $\frac{\gw {S_{\nest {\abs{S_0}}}}}{\gw{S_0}}$ is to be established.
\par
Construction takes place by the addition of $m\geq 1$
arcs (not in $S$) with the respective vertices (that may already
be in $S$). The laws of $S$ (as structure maintained by the $PTA$) 
may or may not allow simple addition of arcs to $S$.
In any general case, however, the weight effect of addition of
$m=\abs{A_{new}}-\abs{A_{old}}\geq 1$
arcs\footnote{$new$ arcs are introduced and $old$ arcs are removed, e.g. If construction by
Nearest Neignbour 
is used then $A_{old}=0$} is $\varDelta A_{\nest{i,m}} 
= A_{new} - A_{old}$.
Thus the total
value of a constructive move is 
$\gw {S_{\nest{i+m}}} = \gw {S_{\nest i}} + \varDelta A_{\nest{i,m}}$.
The relative change during construction is:
\[
R_i =\frac{\gw{ S_{\nest{i+m}}}}{\gw {S_{\nest{i}}}} =
 1 + \frac{\varDelta A_{\nest{i,m}}}
{\gw{S_{\nest i}}}, i\geq 2
\]
Upon completion of $PTA$, the \Q{performance ratio} is
\begin{equation}
\frac{\gw {S_{\nest n}}}{\gw {S_0}} \leq
\sum_{i=1}^n \frac{\varDelta A_{\nest{i,m}}}{\gw{S_{\nest i}}}
;\,\,\,n=\abs{S_0}
\label{PR}
\end{equation}
The $PTA$ has a \emph{minimisation} objective, specifiable
through the use of the average arc weight:
\begin{equation}
\begin{aligned}
\rho_i &=\frac{i}{i+m}R_i = \frac{i}{i+m}\nest{1 + \frac{\varDelta A_{\nest{i,m}}}{\gw{S_{\nest i}}}}\leq 1\\
&\Rightarrow \frac{\varDelta A_{\nest{i,m}}}{\gw{S_{\nest i}}}\leq \frac{m}{i}
\end{aligned}
\label{AvArc}
\end{equation}
Provided \eqref{AvArc} is met throughout (it does not really matter if it is not), a \B{worst-case} construction
behaviour can be infered using \eqref{PR}:
\begin{equation}
1 \leq\frac{\gw{S_{\nest i}}}{\gw {S_0}} \leq m\sum_{i=1}^n 
\frac{1}{i}\leq m\log_2 n
\label{TheLog}
\end{equation}
This is a nice result in its concealment of basic lines of attack to
solving some combinatorial optimisation problems, irrespective of their
value domains. The point is not \Q{$m$ exists!} in \eqref{TheLog}, but
\Q{What of value of $m$ (or $m_i$ for that matter)?} in relation to
\eqref{AvArc} in respect of the TSP. Similar result to \eqref{TheLog}
is given in \cite{142745} and confirmed in \cite{161018}.
\par
\[
***
\]
\par
$\xi$ is a bound by virtue of construction behaviour, and is totally
unrelated to achievable quality of solution by a $PTA$. $\xi$ is
derived as a side-effect of the execution of $PTA$ and is just
a function of solution structure size.
\bibliographystyle{amsplain}
\bibliography{biblio}
\end{document}